# Statistical Inference in a Spatial-Temporal Stochastic Frontier Model

Erniel B. Barrios
*School of Statistics, University of the Philippines Diliman*
Correspondence: ebbarrios@up.edu.ph

John D. Eustaquio
*School of Statistics, University of the Philippines Diliman*

Rouselle F. Lavado
*Philippine Institute for Development Studies*
*(Currently with Asian Development Bank)*

**Abstract**

The stochastic frontier model with heterogeneous technical efficiency explained by exogenous variables is augmented with a spatial-temporal component, a generalization relaxing the panel independence assumption in a panel data. The estimation procedure takes advantage of additivity in the model, computational advantages over maximum likelihood estimation of parameters is exhibited. The spatial-temporal component can improve estimates of technical efficiency in a production frontier that is usually biased downwards. We present a test to verify model assumptions that facilitates estimation of parameters.

**Acknowledgement:** Earlier version of the paper first appeared as Discussion Paper No. 2010-08 in Philippine Institute for Development Studies.

# 1. Introduction

Typical econometric modeling aims to explain the output indicator $y_t$ in terms of determinants, say $x_t$. The error term or the difference between the predicted value $\hat{y}_t$ from the actual value $y_t$ is attributed to other unaccounted determinants, random errors that cannot be accounted by $x_t$ through the specified model, or simply due to model misspecification. The equilibrium assumption also implies that the producer always aims to optimize the output, further implying that the error is a random occurrence. In reality however, some producers may not be efficient enough in utilizing the factors of production available, resulting to the discrepancy between their actual output and the expected optimum output, also labelled as inefficiency. The inefficiency can further be explained by some exogenous factors peculiar to each producer. The result is a stochastic frontier model where the error term in standard econometric specification is decomposed further into those that can be explained by exogenous factors characterizing the producer's inability to maximize the output, and pure error. Stochastic Frontier Analysis (SFA) helps explain producer's efficiency and provide an alternative paradigm to econometric analysis when some assumptions (e.g., equilibrium) fail.

Although the producers by default are the firms, Amos et al (2004), used SFA in studying productivity and technical efficiency of small-scale farmers (producing units). Empirical evidence of the usual assumption that technical efficiency of farmers engaged in mixed crops are generally higher than those propagating only one crop at a time were generated. The notion of a producing unit has been liberally defined in various applications of SFA like in the case of energy efficiency where Huang et al (2015) used the regions of China and Ueasin et al (2015) defined the rice husk biomass power plants as the producing units.

The literature of SFA initially focused on applications in various production setups and estimates of efficiency were noted to vary with model specification and distributional assumptions on the error terms. However, Atilgan (2016), observed that even though varying the specification affects the estimated scores, these scores are found to be highly correlated regardless of the model specification used. The non-robust estimates of efficiency should be addressed through a model specification that describes the production process more realistically.

The literature of statistical modelling on the other hand continues to postulate new models that aim to describe reality in as vivid mathematical abstractions as possible. Spatial and temporal dependence has initially been treated separately. However, as more panel data becomes available and the realization of the potential multiplier effect in information-generation capability of the interaction of space and time, there is a growing interest in spatial-temporal models. A purely spatial model usually has no causative component in it; such models are useful when space-time process has reached temporal equilibrium, or when short-term causal effects are aggregated over a fixed time period, Cressie (1993). A spatial-temporal model then provides a more flexible alternative to postulate a model.

In the simultaneous treatment of space and time, estimation procedures become more complicated. Richardson et al (1992) estimated a spatial linear model with autocorrelated errors, where the spatial and temporal dependencies of the observations yield a general form of the variance-covariance matrix, hence least squares estimate is weighted by elements of the variance-covariance matrix. Iteratively estimated general least squares (EGLS) method was used to sequentially estimate the parameters and the elements of the variance-covariance matrix. More complicated estimation procedures continuously appear in the literature, and the trend is to develop simpler, yet competitive estimation procedures. The backfitting algorithm initially proposed to estimate an additive model (see for example Hastie and Tibshirani, 1990) provides simple alternative to the least squares or maximum likelihood-based estimation procedures. The same algorithm has been used to simplify the estimation procedure for a spatial-temporal model, see for example, Landagan and Barrios(2007).

We propose to augment the stochastic frontier model with spatial-temporal components to allow analysis of panel data even with the violation of the panel independence assumption. The backfitting algorithm is modified in estimating the model.

2. **Stochastic Frontier Analysis**

The extensive literature on SFA has been summarized by Kumbhakar and Lovell (2000). A cross-sectional production frontier model is given by:

$$y_i = f(\boldsymbol{x}_i; \boldsymbol{\beta}) \exp(v_i) TE_i \text{ or } TE_i = \frac{y_i}{f(\boldsymbol{x}_i; \boldsymbol{\beta}) \exp(v_i)} \tag{1}$$

where $y_i$ is the single output of producer $i$, $\boldsymbol{x}_i$ is the vector of inputs to produce $y_i$, $f$ is a parametric function, $TE_i$ is the output-oriented technical efficiency of producer $i$, and $v_i$ is a random error. There is perfect efficiency when $TE = 1$, while inefficiency occurs when $TE < 1$, distance from 1 indicates distance from frontier production level. The shortfall in production environment characterized by $exp(v_i)$ varies across producers. Let $TE_i = \exp(-u_i)$, then the production stochastic frontier model becomes $y_i = f(\boldsymbol{x}_i; \boldsymbol{\beta}) \exp(v_i) \exp(-u_i)$, the last two factors are corresponding error components.

For the parametric function $f$, the literature is dominated by those using the Cobb-Douglas production function family. However, Henderson and Simar (2005) considered a nonparametric specification of $f$, desirable in cases where the modeler is not willing to risk any parametric functional form because of insufficient knowledge about the phenomenon being modeled. A Bayesian formulation of $f$ was considered by Koop and Steel (1999), where contrary to the nonparametric argument, prior knowledge about the efficiency of producers being analyzed are incorporated into the model. Another Bayesian approach on stochastic random frontier model was introduced by Tsionas (2002) which assumes that each producing unit are allowed to have varying coefficients. This model aims to improve the accuracy of the efficiency estimation by permitting heterogeneity across units since Koop and Steel (1999) does not allow heterogeneity. A Bayesian Beta regression approach was used by Senel and Cengiz (2016) to estimate the health system efficiency for all Organization for Economic Co-operation and Development (OECD) member countries. A semiparametric smooth coefficient stochastic frontier model was used by Yao et al (2019) to estimate the productivity of large U.S. firms in a panel data setting as an alternative to using the Cobb-Douglas function.

The model is estimated usually via maximum likelihood estimation (MLE) or its variants. The quantities $v_i, u_i$ and $\boldsymbol{x}_i$, are assumed to be independent and $v_i$ is usually assumed to be normally distributed while $u_i$ is the positive half-normal distribution to ensure that technical efficiency estimates are between zero and one. Other combination of the distribution of $v$ and $u$ include normal-exponential, normal-gamma, and the normal-truncated normal which has a potential to partially alleviate the problem of having not much theoretical reason for the choice of the distributional form of $u$, Coelli et al (2005). Horrace and Parmeter (2015) proposed a Laplace stochastic frontier model where the distribution of inefficiency conditional depends on the composed error is constant for positive values of the composed error and varies for negative values. The nature and relationship between $v$ and $u$ can be enhanced further using mixed model specifications. Green (1990) however, observed that estimates of efficiency vary depending on the distributional assumptions on $v$ and $u$. A model specification that best characterize reality can help improve the robustness property of the estimates.

Since the model is postulated in such a way that efficiency is an upper bound of productive capacity of producers, efficiency estimates are restricted to be biased downwards (i.e., inefficient than they really are). The bias is analyzed by Gijbels et al (1999) in the data envelopment analysis (DEA) estimator which is the set under "lowest" concave monotone function covering all the sample points, for a single input and single output case.

For time series data on the other hand, time-invariant or time-varying technical efficiency were considered. The error assumptions also included fixed and random effects. Heteroskedasticity in $v$ and $u$ was also considered, possibly leading to volatility assumption in technical efficiency. In terms of the parameter estimates, Greene (2005) considered a special case of the random parameters model that produces a random effects model that preserves the central feature of the stochastic frontier model and accommodates heterogeneity. Stochastic frontier models for panel data were postulated with time-invariant technical efficiency assumption, fixed-effects model, random-effects model, or even mixed model. A careful attention on coverage period of the data used in estimation is necessary. Kumbhakar and Lovell (2000) warned that the longer the panel, the less likely it becomes that technology remains constant, a serious violation of the assumption. The learning curve of producers is expected to improve over time and inefficiency realized in the distant past will be less likely to repeat in the future. Battese and Coelli (1992) postulated the following model

$$y_{it} = f(x_{it}; \beta) \, exp(v_{it}) \, exp(-u_{it}) \qquad (2)$$

$$u_{it} = \exp\{-\gamma(t-T)\} u_i \qquad (3)$$

Where $v_{it} \sim NID(o, \sigma_v^2)$ and $u_i \sim NID^+(0, \sigma_u^2)$. Equation (3) characterizes the improving learning curve over time, parameterized by $\gamma$. The likelihood function is easily constructed from the normal and half-normal distributions and maximum likelihood estimators are derived.

One aim of SFA is to explain inefficiency/efficiency in terms of exogenous determinants and Kumbhakar and Lovell (2000) summarized some models to explain inefficiency/efficiency of a producer. Kumbhakar et al (1991) assumed a Cobb-Douglas production function $lny_i = lnf(x_i; \beta) + v_i - u_i$ with $u_i = \gamma' z_i + \epsilon_i$, the exogenous determinant of efficiency is postulated outside the production function, implying additivity of the effect of factors of production and exogenous factors to actual production. Reifschneider and Stevenson (1991) generalized the efficiency equation into $u_i = g(z_i; \gamma) + \epsilon_i$ where $g$ is a nonparametric function. The choice of the best way to analyze the effect of exogenous factors depends on adequacy of the underlying assumption associated with the model. Even a nonlinear regression was used in estimation. The resulting estimates of production efficiencies, however, are expected to vary according to the postulated model.

**3. Spatial-Temporal Stochastic Frontier Model**

Panel data contains information on both the temporal dependencies and the relationship among units at specific point in time. However, if units were selected at one point in time using a probability sampling procedure, oftentimes, the induced sampling distribution characterizes basic independence of the observations. These are the most common models usually postulated for panel data.

At specific time point, dependencies among units can be generated not only by the sampling distribution induced by the selection procedure, but also with those influenced by other units within a specific neighborhood. Several measures of spatial distance have been proposed in the literature of spatial statistics. Depending on the complexity of the model and the problem, simple or complicated measures of spatial distance will be needed. A spatial stochastic frontier analysis (SSFA) approach was proposed by Fusco and Vidoli (2013) where the efficiency was split into three components: spatial lag, decision making unit's (DMU) specificities, and the error term but it does not account for the time-decaying inefficiency that can be observed in a panel data.

In stochastic frontier modelling, several models were proposed given a panel data. Assuming constant factor coefficients over time, Battese and Coelli (1992) postulated a time-decaying inefficiency (improving learning curve).

Over time, the producers get to realize their failure to adopt efficient technologies and correct it soon after which, more efficient production process is applied. Battese and Coelli (1995) further postulated that inefficiencies are function of some exogenous variables and used the maximum likelihood technique in parameter estimation.

Many stochastic frontier models for panel data failed to account for temporal dependencies (improving learning curve of producers) and spatial externalities (adoption of efficiency-enhancing technologies among the producers in a spatial neighborhood) simultaneously. Ignoring this aspect of the information contained in the panel data will result to inadequate differentiation of the producer's efficiency-inducing potentials, hence, may result to inferior estimates of technical efficiency coefficients. Also stated by Chudik et al (2011) and Pesaran and Tosetti (2011) that the spatial model can only accommodate the weak form of cross-sectional dependence (CSD), and thus the spatial based approaches are potentially subject to biases in cases of strong CSD. Mastromarco et al (2016) proposed a way to accommodating both weak and strong CSD in modelling technical efficiency by combining the exogenously driven factor-based approach and an endogenous threshold efficiency regime selection mechanism and makes use of unobserved time-varying factors to account for time-varying technical inefficiency.

A spatial-temporal stochastic frontier model is postulated as

$$lny_{it} = lnf(x_{it}; \beta) + v_{it} - u_{it} \qquad (4)$$

$$v_{it} = \rho v_{it-1} + \psi_{it} \qquad (5)$$

$$u_{it} = \frac{1}{1+exp[-(w_{it}\gamma + z_{it}\phi)]} + \epsilon_{it} \qquad (6)$$

where, the subscript $i$ refer to the producer and $t$ the time period, hence, $y_{it}$ is the output of producer $i$ at time $t$, $x_{it}$ are the factors of production, $v_{it}$ is the autocorrelated pure error, $u_{it}$ are measures of inefficiency, $w_{it}$ are measures of spatial distance, $z_{it}$ are other determinants of inefficiency, $\epsilon_{it}$ and $\psi_{it}$ are a white noise terms $\beta, \gamma, \phi$ and $\rho$ are the corresponding parameters. The production structure is assumed to be constant over time, hence reflected in time-independence of $\beta$. In a reasonably sized panel, production structure is not expected to change since changes may have been brought by significant technological innovations that can be detected only in a much longer panel. The temporal dependence measured by $\rho$ also assumes homogeneity across producers. The short-term dependency in efficiency indexed by $\rho$ is not expected to exhibit structural changes within a short panel. Unlike Battese and Coelli (1995) that specified a non-negative-valued distribution for error terms (hence, a more complicated likelihood function), the logit specification in equation (6) will ensure non-negative predicted values of $u_{it}$ resulting to estimates of technical efficiency bounded above by 1.

A dynamic production parameter in equation (6) may account for the spatial externalities accounted by the spatial indicator but will require more complicated estimation procedure. Equation (5) can also be generalized to higher-order AR process, but the time-adjustment process of inefficiency reduction might be contaminated for much longer autoregressions given a short panel.

The additivity of the models presented in equations (4) to (6) will make estimation via the hybrid backfitting algorithm feasible. The estimation algorithm follows:

(1) Equations (4) and (5) are combined and $-u_{it}$ is ignored to estimate $\beta$ and $\rho$ simultaneously using generalized least squares. Compute the residuals $\hat{u}_{it} = lny_{it} - lnf(x_{it}; \hat{\beta}) - \hat{\rho}e_{it-1}$, this contains information on $\gamma$ and $\phi$. $e_{it-1}$ is the lagged value of the residuals from the fitted model of $v_{it}$.

(2) Given $\hat{u}_{it}$, fit equation (6) as a general linear model to estimate $\gamma$ and $\phi$.

(3) The estimate of technical efficiency is

$$TE_{it} = exp - \left[\frac{1}{1+\exp[-(w_{it}\hat{\gamma}+z_{it}\hat{\phi})]}\right] \qquad (7)$$

The simultaneous estimation of $\beta$ and $\rho$ yield optimality over individual estimation in pure backfitting of an additive model. As noted by Landagan and Barrios (2007), this will not necessitate further iteration of the algorithm.

The inclusion of autoregression in the error of the production function will account for the producers' learning curve while also accounting for the possible cumulative effect of production errors. The spatial externalities that can vary over time and across neighborhoods help characterize efficiency/inefficiency differences among the producers.

## 4. Testing for Constant Temporal Effect Across Space

Estimation of the model is facilitated with the assumption of temporal dependence constant across spatial units. In a short panel data, similar temporal dependence among spatial units are not difficult to assume. However, as the time series length increases, temporal dependence like autocorrelation could possibly vary among spatial units. Non-constant temporal effect across space could lead to estimation procedure discussed in Section 3 with increased residuals from the production function that can still be explained by spatial externalities and would be lumped together into inefficiencies. This assumption of constant temporal effect can be tested using the bootstrap-based inference suggested by Guarte and Barrios (2013). Given the time series in each location/spatial unit, the following hypothesis is tested:

$H_0$: All spatial units have the same autocorrelation coefficient $(\rho)$.

$H_1$: At least one spatial unit have different autocorrelation coefficient.

Using AR-sieve bootstrap, the algorithm for testing the hypothesis of constant temporal effect across space follows:

1. Given the time series $y_{k1}, \ldots, y_{kT}$ for each location $k$. Suppose each with time series can be adequately represented by $AR(p)$ given by

$$lny_{it} = \rho_0 + \rho_1 y_{it-1} + \cdots + \rho_p y_{it-p} + \psi_{it}, t = 1, \ldots, T \qquad (8)$$

with $\psi_{it} \sim NID(0, \sigma_\psi^2)$. Predict $y_{it}$ using the model

$$\ln y_{it} = \hat{\rho}_0 + \hat{\rho}_1 y_{i,t-1} + \cdots + \hat{\rho}_p y_{i,t-p} \qquad (9)$$

where $\hat{\rho}_0, \hat{\rho}_1, \ldots, \hat{\rho}_p$ are maximum likelihood estimates of the corresponding parameters. Let $\rho_p$ be the temporal parameter of interest. Conditional on $(y_{i,t-1}, \ldots, y_{i,t-p})$, the centered residuals have the empirical distribution $e_{it} \sim (0, MSE)$.

2. From the distribution of the centered residuals, generate $k$ bootstrap samples for each location $i$ from a sample of size $m$ say $(e_{i0}^*, \ldots, e_{im}^*)$.

3. Generate $k$ time series for each location $i$, one for each bootstrap sample in Step 2 using the estimated model (6). Each time series $(y_{i1}^*, \ldots, y_{im}^*)$ will be recreated using these two steps:

    a) Initialize $y_{i0}^*, y_{i1}^*, \ldots, y_{i,t-1}^*$.

    b) Compute $y_{it}^*$ using the equation $\ln y_{it} = \hat{\rho}_0 + \hat{\rho}_1 y_{i,t-1} + \cdots + \hat{\rho}_p y_{i,t-p} + e_{it}^*$, t =1, …, m.

4. Estimate the AR(p) model used in Step 1 for each of the simulated time series in step 3.

5. Find the appropriate percentiles for the limits of $\hat{\rho}_p$ to construct a $(1 - \alpha)100\%$ confidence interval.

6. Compute the average of the temporal parameter estimates $\hat{\rho}_p$ in Step 1.

7. Reject the null hypothesis that there is constant temporal effect across location at $\alpha\%$ level of significance if at least one of the intervals fail to contain the computed value in Step 6.

As shown by Guarte and Barrios (2013), this bootstrap procedure yields reliable bootstrap estimates of the temporal parameter if the AR(p) model fitted in Step 1 is correctly selected.

## 5. Testing for Constant Spatial Effect Over Time

Homogeneity of temporal dependence across spatial units is also assumed to facilitate estimation of the additive model. The constant spatial effect over time can also be tested using the bootstrap-based method by Guarte and Barrios (2013). Given the cross-sectional data in each time point, we test the following hypothesis:

$H_O$: All time points have the same spatial effect in the technical efficiency component over time.

$H_1$: At least one time point has a different spatial effect in the technical efficiency component

Given the cross-sectional data $TE_{1t}, \dots, TE_{Nt}$ in each time point $t$. Suppose the regression model

$$-\ln\left[-\frac{1+\ln TE_{it}}{TE_{it}}\right] = w_{it}\gamma_t + \delta_{it}, \; i = 1, \dots, N \qquad (10)$$

where $\delta_{it} \sim (0, \sigma_\delta^2)$. Using the bootstrap for regression analysis, the algorithm for testing the hypothesis of constant spatial effect across time follows:

1. Estimate model (7) and predict $TE_{it}$ using

$$-\ln\left[-\frac{1+\ln TE_{it}}{TE_{it}}\right] = w_{it}\widehat{\gamma_t}, i = 1, \dots, N \qquad (11)$$

   where $\widehat{\gamma_t}$ is a vector of maximum-likelihood estimates of the regression parameters.

2. Compute the average of the estimated spatial parameter of interest in Step 1.

3. Generate $k$ bootstrap samples of $N$ pairs of $(w_{it}, TE_{it})$ from the $N$ pairs of observations on the predictor and response variables in each time point.

4. Estimate the Equation (11) for all bootstrap samples in each time point.

5. Sort the $k$ bootstrap estimates in either ascending or descending order then construct the (1-α)100% confidence interval by finding the appropriate percentiles.

6. Reject the null hypothesis that there is constant spatial effect across time with $(1-\alpha)100\%$ coverage probability if more than if more than $\alpha*100\%$ of the constructed intervals fails to contain the average computed in Step 2.

Resampling cases in regression analysis allows us to compare the spatial effects across time. This bootstrap method is known to be less sensitive to model misspecification according to Guarte and Barrios (2013). If the hypotheses in Sections 5 and 6 are rejected, the estimation procedure presented in Villejo at al (2017) can be used instead.

## 6. Simulation Study

A simulation study is designed to evaluate the power and size of the proposed testing procedure for constant temporal effect across space and constant spatial effect across time in a spatio-temporal stochastic frontier model. For each simulation setting, the hypothesis testing procedure is replicated 2000 times to characterize the proposed test. Table 1 shows the simulation boundaries employed in the study.

**Table 1. Boundaries of Simulation Study**

| | | |
|---|---|---|
| 1 | Sample Size | Small – 50 <br> Medium – 100 <br> Large – 200 |
| 2 | Time Length | Short – 12 <br> Long – 60 |
| 3 | Contribution of the terms in the technical efficiency | a. Equal Contribution of Spatial Effect and Covariates <br> b. Dominating Spatial Effect <br> c. Dominating Covariate |
| 4 | Difference from "true" value of $\rho$ in simulation $[\rho + r\rho]$ | a. $r = 0.0$ <br> b. $r = 0.5$ <br> c. $r = 1.0$ <br> d. $r = 1.5$ |
| 5 | Percent of time points with alternative parameter values | a. 10% <br> b. 20% |
| 6 | Percent difference from "true" value of $\gamma$ in simulation | a. $g = 0.0$ <br> b. $g = 0.5$ <br> c. $g = 1.0$ <br> d. $g = 1.5$ |
| 7 | Percent of spatial points with alternative parameter values | a. 10% <br> b. 20% |

Contribution of each components were considered in the simulation study. This allow us to evaluate whether the test procedure would still provide correct inferences regardless of the contribution of the spatial effect on the model. Power of the test is assessed by adjusting parameter for the temporal/spatial effect by a certain percentage $(r)$ for either 10 or 20% of the spatial/temporal points. The scenarios where $r$ is set to zero was used to determine if the test is correctly sized.

The power of the test for constant temporal effect across spatial units is computed for the scenarios with small or large number of time points where 10% or 20% of the time points have autocorrelation parameter that is different from the rest of the time points. For the computation of the power, the test is replicated 2000 times in every setting where the null hypothesis is indeed false and the proportion of rejection of the null hypothesis is computed for each setting.

In Table 2, even with only 10% of the time points have different values for the autocorrelation, the power of the test is at least 0.90 for cases where the change in the value of the parameter is only 30% regardless of the number of spatial units and time points in the data.. In cases where there is at least a 100% increase or decrease in the value of the autocorrelation in 10% of the time points, the power of the proposed test for testing constant temporal effect across spatial units is at least 0.95.

When 20% of the time points have a different value for the autocorrelation, the power of the test is at least 0.95 except for the case that there are only few spatial units and a short time series. For all cases where the change in the value of the autocorrelation is at 150%, power of the test almost always reached 1.0.

The test is relatively more powerful when sample size is large. Power is also exhibited with smaller sample size provided that the time series is longer.

Table 2 also shows that for short time series, the test is correctly sized. However, as number of spatial units increases, there is a tendency for the test to be incorrectly sized as the length of the time series increases. With more spatial units, there would is larger chance of having extreme values on one of those spatial units that could provide evidence against the null hypothesis. Longer time series increases the chance of relatively high or low values of autocorrelation resulting to incorrect size.

Whether the covariate effect or the spatial effect dominates, power and size of the test are not affected as exhibited in Table 2.

**Table 2. Empirical Size and Power in Testing Constant Temporal Effect in the Error Structure across Locations for Different Simulation Scenarios**

| Simulation Settings | | | Size under $\rho_i = \rho$ | Power at: $\rho_i = \rho + r\rho$ | | | | | |
|---|---|---|---|---|---|---|---|---|---|
| | | | | 10% of time points | | | 20% of time points | | |
| | | | | r=0.3 | r=1.0 | r=1.5 | r=0.3 | r=1.0 | r=1.5 |
| $n = 50$ | $t = 12$ | Equal | 0.042 | **0.903** | **0.941** | **0.995** | **0.925** | **0.950** | **0.999** |
| | | Spatial Effect Dominates | 0.050 | **0.906** | **0.944** | **0.996** | **0.927** | **0.950** | **0.999** |
| | | Covariate Effect Dominates | 0.032 | **0.908** | **0.951** | **0.997** | **0.930** | **0.956** | **0.999** |
| | $t = 60$ | Equal | 0.064 | **0.908** | **0.951** | **0.998** | **0.932** | **0.960** | **0.999** |
| | | Spatial Effect Dominates | 0.062 | **0.914** | **0.955** | **0.998** | **0.945** | **0.968** | **0.999** |
| | | Covariate Effect Dominates | 0.061 | **0.920** | **0.960** | **0.998** | **0.953** | **0.979** | **0.999** |
| $n = 100$ | $t = 12$ | Equal | 0.048 | **0.924** | **0.961** | **0.997** | **0.968** | **0.958** | **0.999** |
| | | Spatial Effect Dominates | 0.042 | **0.931** | **0.961** | **0.997** | **0.970** | **0.959** | **0.999** |
| | | Covariate Effect Dominates | 0.038 | **0.932** | **0.967** | **0.997** | **0.974** | **0.959** | **0.999** |
| | $t = 60$ | Equal | 0.058 | **0.955** | **0.976** | **0.999** | **0.980** | **0.990** | **1.000** |
| | | Spatial Effect Dominates | 0.058 | **0.956** | **0.979** | **0.999** | **0.982** | **0.995** | **1.000** |
| | | Covariate Effect Dominates | 0.055 | **0.963** | **0.981** | **0.999** | **0.984** | **0.995** | **1.000** |
| $n = 200$ | $t = 12$ | Equal | 0.047 | **0.965** | **0.968** | **0.999** | **0.986** | **0.980** | **1.000** |
| | | Spatial Effect Dominates | 0.049 | **0.965** | **0.968** | **0.999** | **0.989** | **0.983** | **0.999** |
| | | Covariate Effect Dominates | 0.039 | **0.966** | **0.972** | **1.000** | **0.992** | **0.983** | **1.000** |
| | $t = 60$ | Equal | 0.053 | **0.969** | **0.984** | **1.000** | **0.995** | **0.997** | **1.000** |
| | | Spatial Effect Dominates | 0.053 | **0.986** | **0.988** | **1.000** | **0.996** | **0.998** | **1.000** |
| | | Covariate Effect Dominates | 0.054 | **0.999** | **0.991** | **1.000** | **0.998** | **0.998** | **1.000** |

Similarly, the test for constant spatial effect in the technical efficiency across time points was evaluated under various simulation scenarios. It can be seen from Table 3 that the proposed test had at least 0.9 power when at least 30% of the spatial units have alternate value of either 30%, 100% or 150% higher than the common spatial effect. This is consistent across the various scenarios with different sizes for the spatial units and different time series lengths. Even for the case that the covariates dominate the technical efficiency term, the power of the test is still nearly 0.90. On the other hand, for the case where only 10% of the spatial units have alternate value, the power of the test is lower at 0.85. For cases with average to large number of spatial units, the power are slightly higher than 0.85. Similar to the result of the simulation for testing for constant temporal effect, the lowest power of the test for this scenario is realized when there are only few spatial units and time series is short. In consideration of the dominant term in the technical efficiency, power of the test is generally higher when the spatial effect dominates the overall technical efficiency while the power is lower for cases wherein the effect of the other factors of efficiency dominates the overall technical efficiency.

The proposed test is correctly-sized for most of the cases where there are small to moderate number of spatial units and when the time series is not very long. The test is fairly robust to which term dominates the technical efficiency equation. Also, test is incorrectly sized when there is a large number of spatial units and with longer time series. The test is correctly sized with short time series and fewer number of spatial units.

**Table 3. Empirical Size and Power in Testing Constant Spatial Effect in the Technical Efficiency across Time Points for Different Simulation Scenarios**

| Simulation Settings | | | Size under $\gamma_t = \gamma$ | Power at: $\gamma_t = \gamma + g\gamma$ | | | | | |
|---|---|---|---|---|---|---|---|---|---|
| | | | | 10% of spatial units | | | 20% of spatial units | | |
| | | | | $g=0.3$ | $g=1.0$ | $g=1.5$ | $g=0.3$ | $g=1.0$ | $g=1.5$ |
| $n = 50$ | $t = 12$ | Equal | 0.047 | **0.855** | **0.885** | **0.990** | **0.890** | **0.950** | **0.990** |
| | | Spatial Effect Dominates | 0.037 | **0.878** | **0.907** | **0.989** | **0.939** | **0.951** | **0.995** |
| | | Covariate Effect Dominates | 0.047 | **0.853** | **0.893** | **0.994** | **0.887** | **0.947** | **0.990** |
| | $t = 60$ | Equal | 0.048 | **0.874** | **0.909** | **0.994** | **0.932** | **0.956** | **0.996** |
| | | Spatial Effect Dominates | 0.047 | **0.872** | **0.915** | **0.996** | **0.944** | **0.956** | **1.000** |
| | | Covariate Effect Dominates | 0.048 | **0.851** | **0.891** | **0.992** | **0.892** | **0.953** | **0.992** |
| $n = 100$ | $t = 12$ | Equal | 0.053 | **0.849** | **0.896** | **0.992** | **0.911** | **0.959** | **0.993** |
| | | Spatial Effect Dominates | 0.052 | **0.870** | **0.902** | **0.996** | **0.914** | **0.971** | **0.995** |
| | | Covariate Effect Dominates | 0.056 | **0.859** | **0.893** | **0.993** | **0.896** | **0.961** | **0.992** |
| | $t = 60$ | Equal | 0.061 | **0.909** | **0.915** | **0.993** | **0.957** | **0.972** | **0.997** |
| | | Spatial Effect Dominates | 0.060 | **0.910** | **0.916** | **0.999** | **0.958** | **0.974** | **1.000** |
| | | Covariate Effect Dominates | 0.061 | **0.906** | **0.908** | **0.997** | **0.954** | **0.973** | **0.996** |
| $n = 200$ | $t = 12$ | Equal | 0.062 | **0.903** | **0.920** | **0.995** | **0.942** | **0.981** | **0.999** |
| | | Spatial Effect Dominates | 0.062 | **0.916** | **0.933** | **0.994** | **0.961** | **0.988** | **1.000** |
| | | Covariate Effect Dominates | 0.066 | **0.901** | **0.921** | **0.992** | **0.945** | **0.986** | **0.998** |
| | $t = 60$ | Equal | 0.066 | **0.932** | **0.933** | **0.998** | **0.962** | **0.990** | **1.000** |
| | | Spatial Effect Dominates | 0.066 | **0.948** | **0.934** | **0.999** | **0.965** | **0.994** | **1.000** |
| | | Covariate Effect Dominates | 0.070 | **0.909** | **0.929** | **0.995** | **0.946** | **0.992** | **1.000** |

## 7. Application

To illustrate the methods, we used the data based on the monitoring of Agrarian Reform Communities by the Philippine Department of Agrarian Reform, collected from the period 2002-2010. The data includes an index of sustainable rural development and various development-enhancing indicators. The panel data on agrarian reform communities (production unit) is analyzed with index of sustainable rural development as output and provision of rural infrastructure and support services as input and determinants of the efficiency equation, respectively. The spatial temporal SFM is labeled as Model 1 and the time-varying decay model of Battese and Coelli (1992) is called Model 2.

The parameter estimates for the factors of production (using the Cobb-Douglas family) are given in Table 4. Similarity in estimates from Models 1 and 2 justifies the assumption of additivity of the model described in equations 4 to 6. Estimates of technical efficiency from Model 1 are generally higher than those coming from Model 2. Model 1 explained further that portion of the residuals from the production function coming from spatial externalities and temporal dependencies are lumped together into inefficiencies in Model 2. The correlation between estimates of technical efficiencies from Model 1 and 2 is 0.5722 indicating that both models identified fairly similar communities to be efficient or inefficient. The efficiency estimates from both models also yield similar correlations with the determinants of efficiency.

The null hypothesis for testing the presence of constant spatial effect across time and presence of constant temporal effect for all location where both not rejected indicating that the spatio-temporal stochastic frontier model fits the data adequately. The non-rejection of the null hypothesis of having constant temporal effect means that the temporal effect in the production does not vary across location. Consequently, the non-rejection of the null hypothesis of having constant spatial effect in technical efficiency across time implies that the there are no significant differences between the efficiency scores yielded by each spatial unit across time assuming all other factors of efficiency constant.

**Table 4. Parameter Estimates**

| Variables | Model 1 | | Model 2 | |
|---|---|---|---|---|
| | Coefficient | p-value | Coefficient | p-value |
| Factors of Production | | | | |
| Log(Index of basic social services) | 0.3542 | 0.0000 | 0.3021 | 0.0000 |
| Log(Index of organizational maturity) | 0.1831 | 0.0000 | 0.1789 | 0.0000 |
| Log(No.Beneficiaries cultivating the land) | 0.0064 | 0.1377 | 0.0038 | 0.0684 |
| Log(No. of Beneficiaries of agrarian reform) | -0.0006 | 0.8830 | 0.0014 | 0.5059 |
| Log(Index of land tenure improvement) | 0.1710 | 0.0000 | 0.1521 | 0.0000 |
| Log(Proportion of credit needs served) | 0.0162 | 0.0000 | 0.0191 | 0.0000 |
| Log(Land area covered by agrarian reform) | 0.0024 | 0.7450 | -0.0115 | 0.0000 |
| Log(Age of the community in the program) | 0.0918 | 0.0353 | 0.0225 | 0.0000 |

## 8. Conclusions

In a stochastic frontier model where specification generally leads to estimates of technical efficiency that is biased downwards, a spatial-temporal component in a panel data model can help improve the estimate. A modified backfitting algorithm that take advantage of the additivity of the models can also facilitate computing especially when large set of factors of production and determinants of inefficiency complicates maximum likelihood estimation in a truncated distribution. The test for constant temporal effect on the efficiency across time and the test for constant spatial effect on the production across spatial units determines whether the proposed model is properly specified and assure that the model is not ill-fitting. The proposed model and the corresponding estimation procedure yields estimates of technical efficiency that are similar to those obtained from some commonly used methods, being able to identify similarly efficient/inefficient producers.

**References:**

Amos T., Chikwendu D., Nmadu J. (2004). Productivity, technical efficiency, and cropping patterns in the savanna zone of Nigeria. *Food, Agriculture and Environment*, 2, 173-176.

Atilgan E. (2016). Stochastic frontier analysis of hospital efficiency: does the model specification matter?. *Journal of Business, Economics and Finance*, 1, 17-26.

Battese G., Coelli T. (1992). Frontier production functions, technical efficiency and panel data with application to paddy rice farmers in India. *Journal of Productivity Analysis,* 3, 153-169.

Battese G., Coelli T.(1995). A model for technical inefficiency effects in a stochastic frontier production function for panel data. *Empirical Economics*, 20, 325-332.

Chudik A., Pesaran M.H., Tosetti E. (2011). Weak and strong cross-section dependence and estimation of large panels. *Econometrics Journal*, 14, 45–90.

Coelli, T. J., Rao, D. S. P., O'Donnell, C. J. and Battese, G. E. (2005). *An introduction to efficiency and productivity analysis 2nd ed*. New York: Springer.

Cressie N.(1993). *Statistics for Spatial Data*. New York: Wiley.

Fusco, E., & Vidoli, F. (2013). Spatial stochastic frontier models: controlling spatial global and local heterogeneity. *International Review of Applied Economics*, 27(5), 679-694.

Gijbels I., Mammen E., Park B, Simar L. (1999). On estimation of monotone and concave frontier functions. *Journal of the American Statistical Association*, 94, 220-228.

Greene W. (1990). A gamma-distributed stochastic frontier model. *Journal of Econometrics*, 46, 141-164.

Greene W.(2005). Reconsidering heterogeneity in panel data estimators of the stochastic frontier model. *Journal of Econometrics*, 2, 269-303.

Guarte J., Barrios E. (2013). Nonparametric hypothesis testing in a spatial-temporal model: A simulation study. *Communications in Statistics – Simulation and Computation*, 42, 153-170.

Hastie T., Tibshirani R. (1990). *Generalized Additive Models*. New York: Chapman and Hall.

Henderson D., Simar L.(2005). A fully nonparametric stochastic frontier model for panel data. Working Paper No. 519, Department of Economics, Binghamton University. http://econ.binghamton.edu/wp05/WP0519.pdf .

Horrace W., Parmeter C. (2015). A Laplace Stochastic Frontier Model. *Econometric Reviews*, DOI: 10.1080/07474938.2015.1059715.

Huang T., Onishi A., Shi F., Morisugi M., Cherry M. (2015). Regional characteristics of industrial energy efficiency in China: application of stochastic frontier analysis method. *Frontiers of Environmental Science & Engineering*, 3, 506–521.

Koop G., Steel M. (1999). Bayesian Analysis of Stochastic Frontier Models, ESE Discussion Papers, Edinburgh School of Economics, University of Edinburgh, https://EconPapers.repec.org/RePEc:edn:esedps:19.

Kumbhakar S., Ghosh S., McGuckin J. (1991). A generalized production frontier approach for estimating determinants of inefficiency in US dairy farms. *Journal of Business and Economic Statistics*, 9, 279-286.

Kumbhakar S., Lovell K. (2000). *Stochastic Frontier Analysis*. United Kingdom: Cambridge University Press.

Landagan O., Barrios E. (2007). An estimation procedure for a spatial-temporal model. *Statistics and Probability Letters*, 77, 401-406.

Mastromarco C., Serlenga L., Shin Y. (2016). Modelling technical efficiency in cross-sectionally dependent stochastic frontier panels. *Journal of Applied Econometrics*, 31, 281–297.

Pesaran M.H., Tosetti E. (2011). Large panels with common factors and spatial correlation. *Journal of Econometrics*, 161, 182–202.

Reifschneider D., Stevenson R. (1991). Systematic departures from the frontier: a framework for the analysis of firm inefficiency. *International Economic Review*, 32, 715-723.

Richardson S., Guihenneuc C., Lasserre V. (1992). Spatial linear models with autocorrelated error structure. *The Statistician*, 41, 539-557.

Senel T., Cengiz M.A.(2016). A bayesian approach for evaluation of determinants of health system efficiency using stochastic frontier analysis and beta regression. *Computational and Mathematical Methods in Medicine*. Volume 2016, Article ID 2801081, 5 pages. doi:10.1155/2016/2801081.

Tsionas, E.G.(2002). Stochastic frontier models with random coefficients. *Journal of Applied Econometrics*, 17, 127–147.

Ueasin N., Liao S., Wongchai A. (2015). The technical efficiency of rice husk power generation in Thailand: comparing data envelopment analysis and stochastic frontier analysis. *Energy Procedia*, 75, 2757 – 2763.

Villejo, S., Barrios, E., Lansangan, J. (2017). Robust estimation of a dynamic spatio-temporal model with structural change. *Journal of Statistical Computation and Simulation*, 87, 506 – 519.

Yao, F., Zhang, F., & Kumbhakar, S. C. (2019). Semiparametric smooth coefficient stochastic frontier model with panel data. *Journal of Business & Economic Statistics*, 37(3), 556-572.